\documentstyle[aps,preprint,epsfig]{revtex}

\def \beq {\begin{equation}}
\def \eeq {\end{equation}}

\begin{document}

\draft

\title {Stability properties and asymptotics for $N$ non-minimally coupled scalar fields cosmology}

\author{L. Brenig$^1$, A. Figueiredo$^2$, 
E. Gunzig$^{1,3}$, T.M. Rocha Filho$^2$
Alberto Saa$^{1,4}$ \footnote{e-mails: 
{lbrenig@ulb.ac.be, annibal@fis.unb.br,  
egunzig@ulb.ac.be, marciano@fis.unb.br, 
asaa@ime.unicamp.br}}}

\address{1)
RggR, Universit\'e Libre de Bruxelles, 
CP 231, 1050 Bruxelles, Belgium.
}

\address{2)
Instituto de F\'\i sica, Universidade de Bra\'\i lia,
70910-900, DF, Brazil.
}

\address{3)
Instituts Internationaux de Chimie et de Physique Solvay, 
CP 231, 1050 Bruxelles, Belgium. 
}

\address{4)
IMECC -- UNICAMP,
C.P. 6065, 13081-970 Campinas, SP, Brazil.}

\maketitle

\begin{abstract}
We consider here the dynamics of some homogeneous and isotropic 
cosmological 
models with $N$ interacting classical scalar fields non-minimally
coupled to the spacetime curvature, as an attempt to generalize
some recent results obtained for one and two scalar fields. 
We show that 
a Lyapunov function can be constructed under certain conditions
for a large class of models, suggesting that  
chaotic behavior is ruled out for them. Typical
solutions tend generically to the empty de Sitter (or Minkowski)
fixed points, and the
previous asymptotic results obtained for the  one field model remain
valid. In particular, we confirm that, for large times and a vanishing
cosmological constant, even in the presence of the extra scalar fields,
the universe tends to an infinite diluted matter dominated era.
\end{abstract}


\section{Introduction}

We have considered recently the homogeneous and isotropic solutions of
the cosmological model described by the action\cite{PRD1}:
\begin{equation}\label{eq1}
    S = \frac{1}{2} \int d^4x \sqrt{-g} \left\{ -(1-\xi \kappa
    \psi^2)\frac{R}{\kappa} + g^{\mu \nu} \partial_\mu \psi
    \partial_\nu \psi - 2 V (\psi)\right\},
\end{equation}
where $\psi$ is the non-minimally coupled scalar field, 
$\kappa = 8 \pi G$, 
$G$ is the Newtonian constant, $R$ the scalar curvature
of spacetime, and the self-interaction potential $V(\psi)$ has the
form:
\begin{equation}\label{eq2}
    V(\psi) = \frac{3 \alpha}{\kappa} \psi^2 - \frac{\Omega}{4}
    \psi^4 - \frac{9 \omega}{\kappa^2},
\end{equation}
with $\alpha = \frac{\kappa m^2}{6}$, $m$ the mass of
the scalar field, $\Omega$ is an arbitrary constant and $-\frac{9
\omega}{\kappa^2}$ is the usual cosmological constant $\Lambda$. 
The nonminimal
coupling term in (\ref{eq1}) is, of course, $\xi R \psi^2$ where $\xi$ is an
arbitrary constant. The results previously obtained
(see \cite{PRD1} for the motivations and references) were worked out
for the case $\xi=1/6$, the so called conformally coupled case,
and they point toward some novel and very interesting dynamical
behavior: superinflation regimes, 
a possible 
avoidance of big-bang and big-crunch
singularities through classical birth of the universe 
from empty Minkowski space, spontaneous entry into and exit from inflation,
and a cosmological history suitable 
for describing quintessence in principle. Through exhaustive numerical
simulations and with some semi-analytical tools, the 3-dimensional 
phase space of the model has been constructed. The existence of a Lyapunov
function for the fixed points was crucial for the study of the
asymptotic behavior\cite{IJTP1} and for precluding the appearance of any 
chaotic regime, confirming and shedding some light on 
 some other previous results in this 
line\cite{chaos}.

The robustness of its predictions must be an essential  
feature of any realistic cosmological model. The study of some
generalizations of the model (\ref{eq1}) is, therefore, mandatory.
In \cite{2fields}, we consider the implications
of the inclusion of a second interacting massless scalar field
 in the dynamics
of homogeneous and isotropic solutions.
The corresponding phase space becomes now 5-dimensional, and 
richer structures could appear. In spite of this, we show that for 
a class of physically 
reasonable
quartic interaction potentials, the neighborhoods of the relevant
fixed points are
unaltered, and, in particular, the asymptotic regimes obtained
for the 1-field case\cite{IJTP1} are preserved in the presence
of an extra massless field. The relaxing of isotropy was the
next performed robustness test of the model. In fact,
it appeared as a corollary of a much more general result.
In \cite{PRD2}, we 
 studied the singularities in the time-evolution of 
homogeneous and anisotropic solutions of cosmological
models described by the action:
\beq
\label{act}
S=\int d^4x \sqrt{-g}\left\{F(\psi)R - \partial_a\psi\partial^a\psi
-2V(\psi) \right\},
\eeq
with general $F(\psi)$ and $V(\psi)$. Such a model has been recently
considered also in \cite{G}. We showed that anisotropic
solutions generically evolve toward a spacetime singularity 
that corresponds to
the hypersurface $F(\psi)=0$ in the phase space. Such singularity
is harmless for isotropic solutions. A second, and different,
type of singularity corresponds to the hypersurface  $F_1(\psi)=0$,
where
\beq
\label{f1}
F_1(\psi) = F(\psi)+\frac{3}{2}\left(F'(\psi)\right)^2,
\eeq
and no solution can avoid it.
 Starobinsky\cite{Starobinski} was the
first to identify the singularity corresponding to the hypersurfaces
$F(\psi)=0$, for the case of conformally coupled anisotropic solutions.
Futamase and co-workers\cite{Futamase} identified both
singularities
in the context of chaotic inflation in $F(\psi)=1-\xi\psi^2$ theories
(See also \cite{s2}).
For this type of coupling, the 
first singularity is always present for $\xi>0$ and the second one for 
$0<\xi<1/6$.
The results of \cite{PRD2} are, however,
more general since  the case of
general $F(\psi)$ is treated and all conclusions
 are based on the analysis of
true geometrical invariants.
The phase space for the model (\ref{act}) is also 5-dimensional, and many new
structures appear. 
The appearance of the singularity of the first type implies 
the instability of
the anisotropic solutions even for the conformally coupled case, 
for instance.
The presence of any amount of anisotropy,
no matter how small, makes the model unstable, challenging
its validity as a realistic cosmological model. 
The asymptotic behavior in the neighborhood
of the fixed points
far from the hypersurface $F(\psi)=0$ is, however, preserved.

We present here a new robustness test for the homogeneous and
isotropic solutions for models of the type (\ref{eq1}).
We consider the case of several interacting scalar fields 
conformally coupled to spacetime curvature. 
Exact
solutions for this case are much more difficult to find, in particular
we could not
identify the heteroclinics and homoclinics as it was done in
\cite{PRD1}. We could, however, get a strong result about
the stability of fixed points suggesting that chaotic
regimes are ruled out. This is
proved for a class of interaction potentials,
through the construction of an explicit 
Lyapunov function. In spite of
the unstable character of anisotropies for models of
the type 
(\ref{eq1}), the results reported here are expected to be valid
in the neighborhood of the Minkowski fixed point, even for the
anisotropic case.

\section{The Lyapunov function}

Let us consider the action for $N$ classical interacting scalar fields
$\psi_1, \ldots, \psi_N$ conformally
coupled to the spacetime curvature:
\begin{equation}\label{eq3}
    S = \frac{1}{2} \int d^4x \sqrt{-g} \left\{ -[1-\frac{\kappa}{6} 
    (\psi_1^2 + \ldots + \psi_N^2)]\frac{R}{\kappa}+\sum_{i=1}^N
    g^{\mu \nu} \partial_\mu \psi_i \partial_\nu \psi_i - 2
    V(\psi_1,\ldots,\psi_N)\right\}
\end{equation}
where the potential $V$ is a polynomial in the $N$ fields
variables $\psi_i$ up to fourth degree.
We consider homogeneous and isotropic solutions corresponding to
Robertson-Walker metrics with flat spatial section
\begin{equation}\label{eq4}
    ds^2=d\tau^2 -a^2(\tau)(dx^2+dy^2+dz^2).
\end{equation}

Variations of the action (\ref{eq1}) with respect to the scalar
fields $\psi_i$ yield $N$ coupled Klein-Gordon equations
\begin{equation}\label{eq5}
    \ddot{\psi}_i + 3 H \dot{\psi}_i-\frac{1}{6} R \psi_i +
    \frac{\partial V}{\partial \psi_i} = 0, 
\end{equation}
where $H=\frac{\dot{a}}{a}$ is the Hubble function and the dot
denotes time derivative.
The variation of $S$ with respect to the metric leads to the Einstein
equations which can be cast, in an analogous way to the  
one field case, in the form
\begin{equation}\label{eq6}
    \frac{\kappa}{2} \sum_{i=1}^N \dot{\psi}_i^2+\kappa V -
    3H^2+\frac{\kappa}{2} H^2 \sum_{i=1}^N \psi_i^2 + H
    \sum_{i=1}^N \psi_i \dot{\psi}_i=0,
\end{equation}
{\em i.e.} the energy constraint, and
\begin{equation}\label{eq7}
    R = -6(\dot{H}+2H^2)=\kappa \left(-4V+\sum_{i=1}^N \psi_i
    \frac{\partial V}{\partial \psi_i}\right),
\end{equation}
the trace equation.
The system of ordinary differential equations (\ref{eq5}) and (\ref{eq7}) 
is defined in a $(2N+1)$-dimensional phase
space. The solutions are, however, confined on the $2N$-dimensional
zero energy hypersurface (\ref{eq6}).

For physical interpretation reasons (the same ones of \cite{PRD1}), 
we assume that the
potential   $V(\psi_1, \ldots,\psi_N)$ contains a mass
contribution in the form of a quadratic term for each field
$\psi_i$, and a homogeneous quartic function $f_4(\psi_1, \ldots,
\psi_N)$ describing the interactions of the $N$
scalar fields (including possible self-interactions), 
and a cosmological constant
\begin{equation}\label{eq8}
    V(\psi_1, \ldots, \psi_N) = \frac{3}{\kappa} \sum_{i=1}^N
    \alpha_i \psi_i^2 + f_4(\psi_1, \ldots, \psi_N)-\frac{9
    \omega}{\kappa^2}
\end{equation}
where ${\alpha_i}=\frac{\kappa}{6} m_i^2$, $m_i$ being the mass
of field $\psi_i$.

We now proceed to show the existence of a Lyapunov function for the
dynamical system (\ref{eq5})-(\ref{eq7}) under some
well defined conditions on the potential (\ref{eq8}).
We focus on the
sub-system of Klein-Gordon equations (\ref{eq5}). By multiplying
each equation by $\psi_i$ and summing on the $N$
fields we get
\begin{equation}\label{eq9}
    \frac{d}{d\tau} \left[ \sum_{i=1}^N \frac{\dot{\psi_i}}{2}
    +V\right]-\frac{1}{6} R \sum_{i=1}^N \psi_i\dot{\psi}_i = -3 H
    \sum_{i=1}^N \dot{\psi}_i^2
\end{equation}
Assuming that $-\frac{1}{6}R\psi_i$ derives from a
potential
\begin{equation}\label{eq10}
    -\frac{1}{6} R \psi_i = \frac{\partial U}{\partial \psi_i},
\end{equation}
we would obtain:
\begin{equation}\label{eq11}
    \frac{d}{d\tau} \left[ \sum_{i=1}^N
    \frac{\dot{\psi}_i^2}{2}+V+U\right] = -3H \sum_{i=1}^N
    \dot{\psi}_i^2.
\end{equation}
If the function between square brackets has a minimum at a
given fixed point, it will be a candidate for a Lyapunov function 
for $H>0$. Equations (\ref{eq5})-(\ref{eq7}) have, obviously,
many fixed points, but we are concerned only with de Sitter ($\omega<0$)
or Minkowski ($\omega=0$) fixed points, for which
$\psi_i=\dot{\psi}=0$ and $H=\pm\sqrt{-3\omega/\kappa}$. We are
restricted, therefore, to $\omega\le 0$ (non-negative cosmological
constants, as in \cite{PRD1,IJTP1}). 
A closer analysis of the energy constraints
(\ref{eq6}) reveals that the system can cross the $2N$-hypersurface
$H=0$ only on a $2N-1$ submanifold of the $2N+1$ original
phase space. In a neighborhood of the fixed point
$(H=\sqrt{-3\omega/\kappa},\psi_i=0)$ where $V$ is non-negative,
the system can reach $H=0$ only at the point where $V$ vanishes,
otherwise the system cannot reach $H=0$. Hence, we will have
for $H>0$
\beq
\frac{d}{d\tau} \left[ \frac{\dot{\psi}_i^2}{2} + V + U\right] \le 0
\eeq
in such a neighborhood 
and, 
provided that $V+U$ has
a minimum at the origin,
\beq
L=\sum_{i=1}^N
\frac{\dot{\psi}_i^2}{2} + V+U  
\eeq
is a Lyapunov function, ensuring, thereby,
the stability of the fixed point $(H=\sqrt{-3\omega/\kappa},\psi_i=0)$.

The relevant question here is the validity of the hypothesis (\ref{eq10}).
It is valid, for instance, if all the fields have the same mass,
 $\alpha_1= \alpha_2=\ldots=\alpha_N$. This is a strong constraint that,
as 
we will see below, can be somehow relaxed.
From equation  (\ref{eq7}), with the potential given by (\ref{eq8}),
we have
\begin{equation}\label{eq13}
    -\frac{1}{6} R \psi_i = \psi_i \sum_{j=1}^N \alpha_j \psi_j^2
    - \frac{6 \omega}{\kappa} \psi_i
\end{equation}
Condition (\ref{eq10}) requires
\beq
\frac{\partial (R \psi_i)}{\partial \psi_k} = \frac{\partial (R
\psi_k)}{\partial \psi_i},
\eeq
implying, from Eq. (\ref{eq7}), that
\beq
\label{couples}
\alpha_k \psi_i \psi_k = \alpha_i \psi_i \psi_k
\eeq
for all couples $(i,k)$ with $i \neq k$. The equal masses
case is, obviously, a particular solution of (\ref{couples}). For this case,
the potential $U$ is readily shown to be
\begin{equation}\label{eq14}
    U(\psi_1,\psi_2,\ldots,\psi_N)=U_0+\frac{\alpha}{4} |\psi|^4 -
    \frac{3 \omega}{\kappa} |\psi|^2
\end{equation}
where $U_0$ is an arbitrary constant and $|\ |$ stands to the usual
Euclidean norm
\beq
|\psi|^2 = \psi_1^2 + \ldots + \psi_N^2.
\eeq
In this case,  the Lyapunov function is given explicitly by
\begin{equation}\label{eq15}
    L = \frac{1}{2} |\dot{\psi}|^2 +
    \frac{3}{\kappa}(\alpha - \omega) |\psi|^2 + f_4 (\psi_1,
    \ldots, \psi_N)-\frac{9 \omega}{\kappa^2} + U_0 + \frac{\alpha
    |\psi|^4}{4}
\end{equation}
for arbitrary homogeneous functions of degree four $f_4$.
In summary, 
the function $L$ has a minimum on the fixed point and
there is a neighborhood 
(the attraction basin) with $H\ge 0$ where 
$L$
has non-positive time derivative. As the system can reach
the hypersurface $H=0$ only in the fixed point, this
provides a stability proof for the fixed point
$\dot{\psi}_i=\psi_i=0$, $H=+\sqrt{{3 |
\omega|}/{\kappa}}$. An analysis of the time-reversed
system reveals that the symmetric fixed point (with $H = -
\sqrt{{3 | \omega|}/{\kappa}}$) must be repulsive. 
The trajectories starting in the vicinity of the last
point will leave this region and some of them will cross the
hypersurface $H = 0$. Some of the crossing
trajectories can be eventually trapped in the half-space $H > 0$ 
and will tend asymptotically toward the
stable fixed point. This phase portrait suggest a regular 
behavior, and it is very reminiscent of the behavior exhibited
by the
solutions of the one-field model, where chaotic regimes were ruled out.

As it was already noted, this reasoning breaks down for $N$ scalar
fields with distinct masses. Nevertheless, we now prove that,
under rather weak conditions relating the masses and the parameters
for some potentials of the form (\ref{eq8}), 
another Lyapunov function exists and
ensures the stability of the fixed point.
We will consider here the special case where the homogeneous
function $f_4$ has the form
(\ref{eq8})
\begin{equation}\label{eq16}
    f_4(\psi_1, \ldots, \psi_N) = - \sum_{l=1}^N \frac{\Omega_l}{4}
    \psi_l^4 + \sum_{k,l=1 (k \neq l)}^N \left(
    \alpha_{lk}\psi_l\psi_k^3 + \frac{\beta_{lk}}{4}
    \psi_l^2\psi_k^2\right)
\end{equation}
with $\beta_{lk} = \beta_{kl}$. 
The Klein-Gordon equations (\ref{eq5}) become
\begin{equation}\label{eq17}
    \ddot{\psi}_i + 3 H \dot{\psi}_i + \frac{6}{\kappa}(\alpha_i -
    \omega) \psi_i + (\alpha_i - \Omega_i) \psi_i^3 + F_i = 0
\end{equation}
where
\begin{equation}\label{eq18}
    F_i = \psi_i \sum_{j=1(j\ne i)}^N
    (\alpha_j+\beta_{ij})\psi^2_j+\sum_{j=1(j\ne i)}^N \alpha_{ij}
    \psi_j^3 + 3 \psi_i^2 \sum_{j=1(j\ne i)}^N\alpha_{ij} \psi_j
\end{equation}
Let us now rescale the field variables by a constant factor 
$    \psi_i = \gamma_i \tilde{\psi}_i$.
Klein-Gordon equations are now
\begin{equation}\label{eq19}
    \ddot{\tilde{\psi_i}} + 3H \dot{\tilde{\psi}}_i +
    \frac{6}{\kappa}(\alpha_i-\omega)\tilde{\psi}_i +
    \gamma_i^2(\alpha_i-\Omega_i){\tilde{\psi}_i}^3+\tilde{F}_i =
    0
\end{equation}
with
\begin{equation}\label{eq20}
\tilde{F}_i = \tilde{\psi}_i {\sum'}_{j=1}^N \gamma_j^2
(\alpha_j+\beta_{ij})\tilde{\psi}_j^2 + \frac{1}{\gamma_i}
{\sum'}_{j=1}^N \gamma_j \tilde{\psi}_j^3
 + 3 \gamma_i \tilde{\psi}_i^2 {\sum'}_{j=1}^N \gamma_j
\alpha_{ij} \tilde{\psi}_j
\end{equation}

We again multiply both sides of (\ref{eq19}) by $\tilde{\psi}_i$
and sum over $i$ leading to
\begin{equation}\label{eq21}
    \frac{d}{d\tau} \left[ \sum_{i=1}^N
    \frac{\dot{\tilde{\psi}}_i^2}{2} + V_1 \right] + \sum_{i=1}^N
    \tilde{F}_i \dot{\tilde{\psi}} = -3 H \sum_{i=1}^N
    \dot{\tilde{\psi}}_i^2
\end{equation}
where
\begin{equation}\label{eq22}
    V_1 = \sum_{i=1}^N \left[ \frac{3}{\kappa} (\alpha_i - \omega)
    \tilde{\psi}_i^2 + \frac{\gamma_i^2}{4} (\alpha_i - \Omega_i)
    \tilde{\psi}_i^4\right]
\end{equation}

Let us determine the conditions under which the term
$\displaystyle{\sum_{i=1}^N \tilde{F}_i \dot{\tilde{\psi_i}}}$ is a
total derivative $\frac{d V_2}{d\tau}$. This requires that
$\tilde{F}_i$ derives from a potential $V_2$, that is
\begin{equation}\label{eq23}
    \frac{\partial \tilde{F}_i}{\partial \tilde{\psi}_k} =
    \frac{\partial \tilde{F}_k}{\partial \tilde{\psi}_i} 
\end{equation}
Conditions (\ref{eq23}) lead to the following algebraic relations
between the masses, the coefficients of the potential $V$ ad the
scaling factor $\gamma_i$,
\begin{equation}\label{eq25}
    \gamma_i^2(\alpha_i+\beta_{ik})=\gamma_k^2(\alpha_k+\beta_{ik}),
\end{equation}
and $    \alpha_{ik} = 0$.
The conditions (\ref{eq25}) provides sign conditions
\begin{equation}\label{eq26}
    \frac{\alpha_i+\beta_{ik}}{\alpha_k+\beta_{ik}} \geq 0
\end{equation}
and homogeneous algebraic linear equations for the $\gamma_i^2$
whose compatibility conditions relate directly the masses
$\alpha_i$ with the interaction coupling constant $\beta_{ik}$.
For $N=2$ (two scalar fields) the condition (\ref{eq25}) are
rather weak
\begin{equation}\label{eq27}
    \frac{\alpha_1+\beta_{12}}{\alpha_2+\beta_{12}} \geq 0.
\end{equation}
For $N=3$, these conditions provide the inequalities (\ref{eq26})
and one strict equality :
\begin{equation}\label{eq28}
    \frac{(\alpha_1+\beta_{12})(\alpha_2+\beta_{23})(\alpha_3+\beta_{31})}
    {(\alpha_1+\beta_{13})(\alpha_2+\beta_{21})(\alpha_3+\beta_{32})}
    = 1.
\end{equation}
It is important to remind that 
$\beta_{ij} = \beta_{ji}$.
For $N=4$, there are four independent conditions
\begin{equation}\label{eq29}
    \begin{array}{l}
    \displaystyle{\frac{(\alpha_1+\beta_{12})(\alpha_2+\beta_{23})(\alpha_3+\beta_{31})}
    {(\alpha_1+\beta_{13})(\alpha_2+\beta_{21})(\alpha_3+\beta_{32})}
    = 1}\\[7mm]
    \displaystyle{\frac{(\alpha_2+\beta_{23})(\alpha_3+\beta_{34})(\alpha_4+\beta_{42})}
    {(\alpha_2+\beta_{24})(\alpha_3+\beta_{32})(\alpha_4+\beta_{43})}
    = 1}\\[7mm]
    \displaystyle{\frac{(\alpha_1+\beta_{12})(\alpha_2+\beta_{24})(\alpha_4+\beta_{41})}
    {(\alpha_1+\beta_{14})(\alpha_2+\beta_{12})(\alpha_4+\beta_{24})}
    = 1}\\[7mm]
    \displaystyle{\frac{(\alpha_1+\beta_{14})(\alpha_3+\beta_{31})(\alpha_4+\beta_{43})}
    {(\alpha_1+\beta_{13})(\alpha_3+\beta_{34})(\alpha_4+\beta_{41})}
    = 1}
    \end{array}
\end{equation}
There are four conditions and eight parameters $\alpha_i,
\beta_{ij}$. More generally, the total number of parameters
$\alpha_i, \beta_{ij}$ is $\frac{N^2}{2}$ and the total number of
conditions (strict equalities) is $\frac{N(N-1)(N-2)}{6}$. We,
thus, can determine the maximum number of interacting scalar
fields above which the above conditions are overdetermined
\beq
\frac{N^2}{2} < \frac{N(N-1)(N-2)}{6}
\eeq
which yields the solution $N>5$.
Finally, the explicit expression of the Lyapunov
function in the original variables $\psi_i$ is
\begin{eqnarray}
L &=& \frac{1}{2} \sum_{i=1}^N \frac{1}{\gamma_i^2} \dot{\psi}_i^2
+ \frac{3}{\kappa} \sum_{i=1}^N \left(
\frac{\alpha_i-\omega}{\gamma_i^2}\right) \psi_i^2 + \frac{1}{4}
\sum_{i=1}^N \left( \frac{\alpha_i-\omega_i}{\gamma_i^2}\right)
\psi_i^4 \nonumber \\[3mm]
&\,& + \frac{1}{4} \sum_{i=1}^N \sum_{j=1 (j \neq i)}N \left(
\frac{\alpha_i-\beta_{ij}}{\gamma_j^2}\right) \psi_i^2 \psi_k^2
\label{eq30}
\end{eqnarray}
with the conditions (\ref{eq25}) over the parameters $\alpha_i,
\beta_{ij}$, and $\gamma_i^2$.

Again, this function vanishes at the fixed points and on the
$H$-axis. It is non-negative definite in a finite domain of the
phase space around that axis and its time derivative is given by
\beq
\frac{dL}{d\tau}=-3 H\sum_{i=1}^N \frac{1}{\gamma_i^2}
\dot{\psi}_i^2
\eeq
which is non-positive in the half-space $H \ge 0$.
This proves the stability of the fixed point $\dot{\psi}_i =
\psi_i = 0$, $H = \sqrt{\frac{3 |
\omega|}{\kappa}}$. All the solutions tend to this point or go to
infinity.
Furthermore, both Lyapunov functions may be global, {\rm i.e.}
\beq
\lim_{|\psi|\to \infty} L = + \infty
\eeq
when the parameters satisfy some inequalities. In that case, the
Lyapunov stability theory shows that solutions are bounded and
accumulate on one of the fixed points, depending on the initial
conditions.

\section{Conclusion}

We have shown under which conditions some $N$ non-minimally
scalar fields 
homogeneous and isotropic cosmological models admit a Lyapunov
function for their de Sitter (or Minkowski) fixed point. The physical
interpretation of such conditions is still unclear.

\acknowledgments

The authors would like to acknowledge financial support from the
EEC contract \#HPHA-CT-2000-00015 from the OLAM-Fondation pour la
Recherche Fondamentale (Brussels) and from CNPq (Brazil).

\end{document}